# Repositories with Direct Representation

Robert B. Allen

Yonsei University

rballen@yonsei.ac.kr

**Abstract:**
A new generation of digital repositories could be based on direct representation of the contents with rich semantics and models rather than be collections of documents.  The contents of such repositories would be highly structured which should help users to focus on meaningful relationships of the contents.  These repositories would implement earlier proposals for model-oriented information organization by extending current work on ontologies to cover state changes, instances, and scenarios.  They could also apply other approaches such as object-oriented design and frame semantics.  In addition to semantics, the representation needs to allow for discourse and repository knowledge-support services and policies.  For instance, the knowledge base would need to be systematically updated as new findings and theories reshape it.

## 1. DIRECT REPRESENTATION

There is an increasing number of non-traditional collections which are different from traditional collections with discrete documents.  These collections are often composed of widely disparate types of materials which overlap considerably in content.  Such collections range from digital humanities, to architecture (e.g., the Façade digital preservation project), and to disaster archives (e.g., CEISMIC).  Even collections with traditional documents are being transformed as we can now readily access the full text of those documents.

We propose direct representation of the contents of repositories.  This would involve representation of entities and models of phenomena.  We use the phrase direct representation to emphasize that we are describing phenomena directly rather than describing documents with metadata.  This is analogous to direct manipulation which refers to user interfaces controlled by pointing devices rather than command languages.

In order to provide access to the contents of documents, we need rich description of the contents beyond the traditional view of documents as linear and monolithic to allow better access to and use of the contents of the documents.  In addition, the rich structure would provide a solid foundation for document processing tasks such as text mining.  There would be many other advantages to this approach.  The goal for most users is not to search for, identify, select, and obtain documents.  Rather, the goal for many if not most users to is to find information.

Over the past 50 years, there has been extensive research in representing semantics.  Recently, semantics has received greater attention with discussion of the Semantic Web, Linked Data, and ontologies.  We believe it is helpful to go beyond loosely linked data and ontologies to consider models which have rich semantics and which allow detailed description of instances, representation of complex entities, and representation of changes through time.  The models we consider include processes, which model both domains and instances.

Direct representation builds on a large body of research in areas such as knowledge representation, hypertext theory, discourse, narrative, explanations, psychology, information organization, process and service specification, text processing and generation, simulation, and programming-languages.

On one hand, it seems natural to build representations from the rich semantics of ontologies.  On the other hand applications ranging from history to scientific research seem more amendable to sequential presentations of explanations and narrative.  Thus, we consider ways to combine these two approaches.  Other desirable properties of models include stability, coverage, interoperability/composability, and generality.  While we believe in the value of structured knowledge representation, we de-emphasize the ability do complex inference and to use reasoners.

Section 2 considers ontologies and some of their limitations. Section 3 examines issues around state changes and causation. Section 4 considers programming languages, especially object-oriented languages. Section 5 focuses on discourse. Section 6 looks at additional features needed for direct representation repositories.  Section 7 discusses developing a direct representation repository for scientific research reports.  Section 8 discusses direct repositories of historical materials. Section 9 is the conclusion.

## 2. ONTOLOGIES

Ontologies link entities by relationships.  Typically, they provide coverage of entities and relationships for the topics associated with a given task or in some cases for the broader domain from which that task is drawn.  This means that many ontologies are ad hoc and do not systematically consider the definitions of the domain nor do they evaluate the completeness of the coverage of that domain.

There is also considerable variety in the semantic relationships on which the domain ontologies are based.  Upper ontologies attempt to provide consistent frameworks of entities.  Several upper ontologies have been proposed.  The Basic Formal Ontology (BFO) is a realist ontology roughly based on Aristotelian principles.  It is particularly important in biology and medicine and includes frameworks both for revising the upper ontology and for coordinating domain ontologies based on it [28]. We have explored applying the BFO to direct repositories [11, 15] beyond its current application areas to scientific communication and for historical materials.

Most ontologies model domains rather than instances of objects. The BFO allows descriptions of instances but this aspect of the approach has not been well studied. Indeed, it remains unclear when an instance is distinct enough to be identified as a separate entity type. In addition, descriptions of instances are distinct from descriptions of domain-level entities because the instances need to be able to include specific states.

Real-world entities are generally complex. An organism could be represented as a specific entity but it is also made up of many parts and we need ways to talk about the interaction of those parts. Thus, an entire organism may be said to have a disease but that disease may be due to a specific mechanism within its cells. Indeed, the multiple layers of complex entities contribute to confusion in representation due to multiple inheritance and to exceptions to causal rules.

When we instantiate a complex entity, that is when we generate a description for an object in the real world, there will undoubtedly be aspects of the complex entity of which we have no knowledge. For example, we would not know the composition of all cells of a person when we first establish a medical record for that person. Similarly in casual descriptions we make inferences based on expectations without any detailed knowledge. The BFO partially addresses this problem by considering entities at different levels of granularity but we believe that difficulties remain.

Another problem in describing the interaction of entities in the real world is that they may be part of ad hoc collections of objects whose overall interaction is unpredictable although individual exchanges may be comprehensible. The situation may be comparable to the difficulty of modeling fluid dynamics. Thus, we need ways to describe complex interactions even if we cannot model them directly. Along the same lines, when we consider the interactions of several people their complex interactions may be captured in a narrative. We need representations for aspects of the narrative such as the scenario [21] in which the interaction takes place.

## 3. PROCESSES

### 3.1 Processes and States

Entity instances change. In natural language we describe those changes with verbs. We also need to allow such changes in highly-structured approaches. However, modeling of such changes are both controversial and complex for natural language (e.g., [20]) and for ontologies (e.g., [28]) alike.

One of the main features of BFO is to allow both 3D and 4D representations of entities [18]. The 3D entities are known as Continuants. The 4D entities are known as Occurrents and they include Processes which are ongoing activities such as running, eating, and talking. This turns out to be confusing because in other frameworks Processes are sequences of different activities.

Upper ontologies differ greatly in how they handle states, state changes, and events. The BFO does not define states but we may consider them roughly as a relationship between a Process and an Independent Continuant. Thus, we may say that a person is in a state of running, eating, or talking. Of course, states may change. These changes are central to our discussion of causation and programming languages below. A collection of state changes such as changing speed while running may be considered a Process Aggregate and perhaps it would be described with a Process Profile.

### 3.2 Processes Chains

Beyond changes in specific states, there are many ways in which sequences of activities can be described. Most typically there are sequences in which state changes among different, but related entities are described. Consider the process of applying for a driver's license or the processes specified in Business Process Engineering. We may describe such sequences of processes as Process Chains.

Some types of Process Chains occur frequently with other Process Chains are unique. Some are directly connected by a causal flow, we could say by a cascade of events. Other Process Chains are externally controlled. Moreover, some Process Chains have branching and even loops. In addition they may be specified at different levels of abstraction.

Different ways of describing this range of alternatives leads to ambiguity. [15] has described Procedures which are sequences that are specified as part of a legal system, as dependent continuants. FrameNet [23] gives the example of a Criminal Process which includes various paths through the criminal justice system. The BFO describes Histories as a collection of different processes associated with a specific entity (e.g., a person's life history). Narratives are sequences of events which typically describe an arc of changes of one or more entities. There is also the important case of Workflows which are closely related to computer programs.

### 3.3 Frame Semantics and FrameNet

Natural language may be considered as a modeling language. However, natural language is so full of ambiguity, that it raises many challenges. The WordNet project has attempted to describe the semantic relationship of word meanings but it moved from describing words to describing ad hoc units of related meaning called synsets. Similarly, the nature and relationships among verbs has been an active topic in linguistics.

Frames are a common knowledge representation structure. They provide an alternative structure to RDF triples. One way that frames have been applied is in the linguistic theory of frame semantics. Frame semantics suggests that meaning arises not from individual words or from a traditional sense of syntax. Rather, meaning arises from the instantiation of frames that describe certain scenarios which include grammatical units such as agents.

As an application of frame semantics, the FrameNet project attempted to build a linguistic resource with frames extracted examples of actual language use. Many of the FrameNet frames implement state changes. For instance the release frame describes the process of being released from a state of captivity. Thus, FrameNet is a major linguistics approaches to verbs. [9] used frames as part of a structured way to implement state changes as object-oriented models for histories.

## 3.4 Causation

The notion of causation is central to many of the many of the issues addressed in this paper. However, there is a great deal of confusion about causation. We believe that a rich model like an upper ontology will go a long way to disentangling interpretation of causation. It can be particularly difficult to disentangle causation after the fact for aggregates of entities, complex entities, or complex scenarios.

[14] proposed that a causal mechanism may be described as an entity in a given state $X_1$ causing a state change in entity Y. That can be written as: $X_1 \rightarrow Y_1, Y_2$. We also note that there are some additional cases such as causes creating or destroying entities.

To create a triple from the causal statement which would be analogous to RDF triples, we need to treat the state as a unit. It is difficult to see how to accomplish that directly with BFO which does not consider states or events as entities. There are other upper ontologies which do support events, but those have limitations in other areas. While we believe that causes are different from ontological relationships, causal statements may still be based on the same upper ontology used for the ontological relationships.

## 4. PROGRAMMING LANGUAGES

Programming languages are a broad approach to modeling. A program is set of mechanisms built on state changes. Unlike ontologies, programming languages implement state changes directly. Remarkably, there has been little attempt to reconcile programming languages with ontologies.

### 4.1 Object-Oriented Models and Languages

The ideal of object-oriented modeling is to identify objects in an environment and then show how they can interact to produce certain outcomes. Object-oriented models were developed first for simulations and have been applied primarily to IT and business applications.

The term "object-oriented" is used in many ways. Object-oriented may simply refer to the use of entities and classes without any interaction among the classes at all. These are not programs. In object-oriented programs there is interaction between the classes (objects) by message passing. Object-oriented programs also use encapsulation which means that both the variables and code necessary to act on messages are internal to each object. Object-oriented languages also typically include inheritance, which is viewed as a way to reduce the complexity of the code.

### 4.2 Relationships between Object-Oriented Programs and Ontologies

The separation of continuants and occurrents in upper ontologies seems closely related to separation of data and methods in object-oriented design. Similarly the extensive use of inheritance in both paradigms suggests there are likely to be other parallels.

Programming languages apply data typing to manage the use of different classes of variables. Most often these data types are types of numerical values such as integers, floats, and doubles. For instance, a compiler might enforce a rule that specific types of variables cannot be combined with specific operators. Depending on design of the language character variables might not be able to be combined with floats. Data typing could also determine what variables are passed to functions or in object-oriented paradigms to other objects. Moreover, it is possible to use semantic types instead of data types.

The causal mechanism described above in Section 3.4 can now be re-written as: $Y_2 = f(X_1, Y_1)$. Where $f(X_1, Y_1)$ is a function or method in the object-oriented programming sense. This formulation also allows including other arguments as needed. Thus, where the = represents assignment to a new value through causation we could read this as $Y_2$ is caused by the interaction of $X_1$ and $Y_1$ For example, we could say: (boiling liquid) (is caused) (by applying energy to non-boiling liquid).

### 4.3 Object-Oriented Programs and Ontologies

Based on these general principles of object orientation, a great many object-oriented computer languages have been developed. Familiar object-oriented languages such as Java and C++ are usually not regarded as prototypical while the languages derived from Smalltalk are considered prototypical. [11] explored the Slate programming language as a framework for object-oriented ontologies. In addition to message passing and encapsulation, Slate has multiple inheritance. It uses prototypes as an approach to inheritance and instantiation. Furthermore, Slate is interpreted. These attributes give Slate specific strengths and weaknesses as a tool for semantic modeling. Other related languages would have other strengths and weaknesses. [11] interwove cycles state-change updates in the ontology updates using Java with validation of the those updates with Jena.

## 5. DISCOURSE

Discourse is the way semantics are used in communication. Effective discourse is essential for supporting user access to a direct representation repository. Discourse is often contrasted to semantics although most previous studies of discourse structure ignore the semantics. However, because discourse is so closely connected to semantics, it must also be interwoven with the structure of the semantics as based on the ontology.

Claims are types of statements about entities that are asserted to belong in the semantic model. For example, they could be statements about the structure of the domain ontology, about instances, or about relationships and causation. We view evidence as a type of claim but as we have seen before, the model with which the evidence is interpreted is also crucial

Discourse goes beyond applying simple labels to passages of discourse. For instance, we may identify a concession in an argument. But, we should also consider how several labels fit together into a macro units [5] such as an explanation, narrative, or argument.

Explanation and narrative are closely related. Explanation describes causal relationships and threads while narrative follows the path of the causal thread. More specifically, there are two senses of narrative. The first is full-blown story telling which includes dramatic elements. The second is simply a retelling of the state changes and conveying the causal processes by showing rather than explaining them.

Scholarly research often involves comparing evidence and models. Probably the best known macro-unit for argumentation is from Toulmin. This macro unit separates an argument into discourse components. However, applying it to transcripts of actual court arguments has proven complex [24] because the issues are too interwoven. Once again, we believe that rich modeling will help keep track of alterative positions.

# 6. SUPPORTING THE REPOSITORY

Management of the repository must go beyond policies for specifying the structure of the knowledgebase and associated discourse. It must provide frameworks and policies for long-term utility and continuity.

## 6.1 Descriptive Programs

As we have suggested earlier the contents of the repository should be increasingly structured. They should become more like a computer program. Taken to the conclusion, the repository could be composed not of traditional documents but of highly-structured descriptive programs.

There could be multiple threads through individual programs and through collections of programs so that there would be multiple paths through them. Ultimately, the programs might be disaggregated so that ad hoc hypertextual threads could be created as needed [6, 16].

## 6.2 Repository Features, Services, and Policies

While rich semantics and discourse provide descriptions for events, a repository may benefit from specific types of annotations. Annotations might highlight gaps in knowledge [Swales], inconsistences, and alternative models. There should also be programs for adding annotations as well as programs for validating the knowledgebase. And, there needs to be extensive version management.

We may identify use cases for different types of users. There would be Casual Users, Scholars, Authors, and Repository Managers. Services such as searches could be tailored for each of these groups. There could also be personalization of paths through the repository based on the interests and sophistication of the users (e.g., [5]). We could also provide support for different cultural and theoretical perspectives For instance, we could develop behaviorist approaches which do not allow representations of mental acts.

Finally, we need a realistic assessment of practicalities such as business models. Similarly we need to determine the burden this approach might place on authors. Authors are already burdened and will be unlikely to make substantial additional effort to satisfy onerous repository requirements.

## 6.3 Unified versus Coordinated Repositories

Scientific research is part of history. But science can also help to explain history. Publishing itself is affected by both science and history while also affecting them. Thus, ultimately, our goal is a conceptually unified approach across all areas of knowledge. We believe that such a framework would also support personalization. However, such a unification of conceptual frameworks must be developed in stages. At least until repositories are fully conceptually integrated there could be a new set of tools comparable to OAI to support coordinated federation of repositories.

# 7. EXAMPLE: HIGHLY-STRUCTURED SCIENTIFIC RESEARCH REPORTS

In the following two sections we consider how collections might be created of scientific research reports and of historical materials. Science is about generalization while history is about instances. Nonetheless, we view history and science as related because they both involve causal processes and modeling.

Although it is widely recognized that science is based on modeling there has been little attempt to systematically incorporate models into scientific communication. This is related to the insight of daWaard [17] about the relationship of scientific communication to narrative and to our previous work [1, 3, 7, 8, 14].

Scientific research reports have become increasingly structured. We propose take that trend to the next level by placing structure above text. We expect to be able to reconstruct an approximation of the text version of research reports from highly-structured research reports, so a program which operates on the research report structures would become the primary deliverable (see Section 6.1).

Highly-structured research reports would need to coordinate two distinct types of processes and structures: the phenomenon being described and the research procedure and results. The description of the phenomenon under investigation would be based on causal processes such as those described in Sections 3.4 and 4.2. Research workflows would be also be based on those sections as well as on object-oriented models for business process engineering (e.g., [22]) and more recent workflow specification models [19, 30]. The latter would also incorporate approaches to research methods [3, 11] and to the specification of data sets. Indeed, research data could be included directly as part of the structured document by extending and applying techniques for describing research measurements from [27].

## 8. DIRECT REPOSITORES FOR HISTORICAL MATERIALS

Like science, history is concerned with processes, causation, generalization, and modeling [2, 4, 7, 8, 13, 25]. Thus, we should be able to describe history with the structured frameworks developed in this paper.

We have been working on specifications for a digital library of the large number of digitized historical newspapers that are now available online. Because that content is so varied and complex, we concluded that the goal should not be to index individual newspaper articles but should be to develop models of the community that is the subject of the newspaper and to use the models to organize the content. Moreover, the community models might incorporate other materials such as census records, oral histories, diaries and museum artifacts.

At the most basic level we could model aspects of the physical infrastructure of the community. Other types of modeling such as models of laws, institutions, and the economy are increasingly complex. Moreover, change in all dimensions is a constant feature. Ultimately, we need to incorporate technological changes as well as the diffusion of those innovations, changes in culture, and even changes in the style of news reporting. There are even greater challenges in modeling people's motivations and in describing social obligations [26].

Once we have a structured repository for a community, we would need to support user interaction with it. Perhaps a narrative could be constructed for the contents of the repository applying techniques such as those developed for automated summarization. A user might explore the contents along threads which follow the sections of the newspaper [12]. Potentially the rich knowledge structure could greatly strengthen the text summarization. But, the text generation would need to be robust to manage the many gaps. While human historians can overcome these challenges [10, 11], it will remain very difficult for an automated system.

## 9. CONCLUSION

With increasing amounts of full text available and with the need to represent a wide range of non-textual materials we believe that direct representation based on rich semantics will be a useful alternative to traditional document-based repositories. Our approach adapts knowledge-organization systems which integrate these components.